\documentclass[prb,aps,twocolumn,showpacs]{revtex4}
\usepackage{graphicx,latexsym}
\usepackage{dcolumn}
\usepackage{amsmath,amssymb,epsf,bm}
\begin{document}
\title{Non-local effect of a varying in space Zeeman field on the supercurrent and the helix state in a spin-orbit-coupled s-wave superconductor}
\author{A.~G. Mal'shukov}
\affiliation{Institute of Spectroscopy, Russian Academy of Sciences,
142190, Troitsk, Moscow, Russia}
\begin{abstract}
A weak parallel Zeeman field combined with the spin-orbit coupling can induce the supercurrent in an s-wave two-dimensional superconductor. At the same time, the thermodynamically equilibrium state of such a system  is characterized by the helix phase where the order parameter varies in space as $\exp(i\mathbf{Qr})$. In this  state the electric current that is induced by the Zeeman interaction is exactly counterbalanced by the current produced by the gradient of the order-parameter. We studied the interplay of the helix state and magnetoelectric current in the case of a varying in space Zeeman field, as it might be realized in hybrid heterostructures with magnetic and superconducting layers. The theoretical analysis was based on Usadel equations for Green functions in a dirty  superconductor. It is shown that even a weak inhomogeneity produces a strong long-range effect on the magnetoelectric current and the order-parameter phase. Consequently, depending on the macroscopic shape of such an inhomogeneity, either the helix state with the zero supercurrent, or a locally uniform state with the finite supercurrent are realized. A mixture of these two extreme situations is also possible. It is also shown that the current can be induced at a large distance from a ferromagnetic island embedded into a superconductor. Quantum effects associated with the magnetoelectric effect are briefly discussed for multiply connected systems. The theory proposes a new point of view on interplay of the magnetoelectric effect and helix phase in spin-orbit coupled superconductors. It also suggests an interesting method allowing to couple superconducting and magnetic circuits.
\end{abstract}
\maketitle

\section{Introduction}

Interplay of superconductivity and magnetism results in a variety of striking physical phenomena that can be observed in very different systems ranging from heavy-fermion noncentrosymmetric superconductors to various hybrid systems, which incorporate superconducting and magnetic components. Some of these solids represent a special group of topological superconductors \cite{topoSup,Majorana} having a great potential for applications in a fault tolerant quantum computing. On the other hand, even systems of a trivial topology can exhibit quite unusual physical characteristics that are determined by fundamental quantum laws, which control superconducting systems. In particular, it has been predicted that the Zeeman interaction of electron's spins with static exchange, or external magnetic fields, in combination with the spin-orbit coupling  (SOI) can induce the supercurrent in a two-dimensional (2D) superconductor having a spatially uniform order-parameter. \cite{Edelstein,Yip} A similar phenomenon has also been predicted for Josephson junctions. \cite{ISHE,phijunction} A possibility to induce a macroscopic current in a thermodynamically equilibrium superconductor places these phenomena in the same category as the Meissner effect. Besides this magnetoelectric effect, it has also been shown that the interplay of a weak parallel Zeeman field and SOI gives rise to a helical state in a 2D superconductor,  \cite{Edelstein,Samokhin,Dimitrova} where Cooper pairs are composed of particles with a shifted in $k$-space center of gravity. In this state the superconducting order-parameter  varies in space as $\Delta=|\Delta|\exp(i\mathbf{Q}\mathbf{r})$, where $\mathbf{Q}$ is the helix wave-vector. It turned out that the equilibrium supercurrent in such a helical state is absent, \cite{Edelstein,Samokhin} because the currents induced by the spatially uniform Zeeman field $\mathbf{Z}$ and by the order-parameter phase gradient cancel each other \cite{Edelstein}. On the other hand, a question arises as to whether a finite current can be induced by a varying in space $\mathbf{Z}$. Indeed, in the spin-orbit coupled superconductor $\mathbf{Z}$ produces the major effect on the phase of the order parameter. At the same time, in superconductors the static perturbations of the order-parameter phase propagate over long distances. Therefore, it is reasonable to expect that the long-range effect of  spatial variations of $\mathbf{Z}$ on the phase of the order-parameter and, hence, on the current $\mathbf{J}$ will be strong.

In order to give a better insight into the interplay of the magnetoelectric current and helix phase, let us consider a bit simplified, but physically transparent interpretation of this phenomenon. Let us assume that SOI is represented by the Rashba \cite{Rashba} interaction $\alpha(k_x\sigma_y-k_y\sigma_x)$, where $k_i$ and $\sigma_i$ are the wave-vectors and Pauli matrices, respectively. The total spin-dependent interaction is represented by the sum of SOI and the Zeeman interaction. Hence, two electron energy bands are given by $E_{\eta}(\mathbf{k})=k^2/2m +\eta \sqrt{(\alpha k_x+Z_y)^2+(\alpha k_y-Z_x)^2}$, where $\eta=\pm$. By assuming $\alpha k_F \gg Z$, where $k_F$ is the Fermi wave-vector, the square root may be expanded  up to linear  in  $Z$ terms. As a result, by neglecting small terms $\sim Z^2/\mu^2, Z\alpha k_F/\mu^2$, where $\mu$ is the chemical potential, the energies $E_+$ and $E_{-}$ can be represented as $E_{\eta}= (\mathbf{k}+\eta \mathbf{A})^2/2m+\eta\alpha |\mathbf{k}+\eta \mathbf{A}|$, where $\mathbf{A}=m(\mathbf{Z}\times\mathbf{e}_z)/k_F$ and $\mathbf{e}_z$ is the unit vector parallel to the $z$-axis. Hence, $\mathbf{A}$ represents a shift of electron's bands in $k$-space, so that the bands with "+" and "-" helicities are shifted in opposite directions. Since, due to the Rashba interaction, the energies of electrons with opposite helicities $\eta$ are split, the occupancies of these bands are different. Therefore, the average helicity $\langle\eta\rangle=(n_+-n_-)/(n_++n_-)\neq 0$, where $n_{\pm}$ are respective electron occupancies. At $\mu\gg\alpha k_F\gg |\Delta|$, where $\Delta$ is the superconducting order-parameter we obtain $\langle\eta\rangle=2\alpha m/ k_F$. As long as we are interested in  effects that are linear with respect to $\mathbf{A}$,  it is reasonable to  approximate $\mathbf{A}$ by its average value. By substituting $\eta \rightarrow \langle\eta\rangle$  we obtain  $\langle \mathbf{A}\rangle \equiv \langle\eta\rangle \mathbf{A}=2\alpha m^2 (\mathbf{Z}\times\mathbf{e}_z)/k_F^2$. At constant $\langle \mathbf{A}\rangle$ one may gauge out this field from the Hamiltonian by the transformation $\psi \rightarrow \exp(-i\langle\mathbf{A}\rangle \mathbf{r})\psi$, where $\psi$ is the field operator. This means that the order parameter varies in space as $\exp(-2i\langle\mathbf{A}\rangle \mathbf{r})$, in accordance with Refs. \onlinecite{Edelstein,Samokhin,Dimitrova}. It is surprising that quite simple arguments result in the same expression $Q=2\langle A\rangle=4\alpha m^2 (\mathbf{Z}\times\mathbf{e}_z)/k_F^2$ for the helix wavevector,  as a rigorous theory \cite{Dimitrova} (at $\alpha k_F \ll |\Delta|$ the expression for $Q$ is different \cite{Edelstein}). Since $\langle\mathbf{A}\rangle$ is absent in the transformed Hamiltonian, the electric current, as expected, is zero. On the other hand, let us assume that the vector $\langle\mathbf{A}\rangle$ varies in the direction that is perpendicular to $\langle\mathbf{A}\rangle$. Then, $\langle\mathbf{A}\rangle$ looks as a transverse vector-potential that could be associated with a magnetic field in the $z$-direction. Hence, a finite supercurrent is induced due to the Meissner effect. In contrast, if $\langle\mathbf{A}\rangle$ is a longitudinal vector, it can be gauged out and does not produce any current. It results, instead, in helix oscillations of $\Delta$.

A goal of this work is to construct a  theory of the magnetoelectric effect produced by a varying in space Zeeman field that is parallel to a 2D superconductor. Such a field can appear in hybrid systems due to the exchange interaction caused by a close contact of superconducting and magnetic systems, for example, in the case of a magnetic insulator island deposited onto a superconducting film. Such sort of systems have been discussed in Refs.\onlinecite{topoSup,Majorana}. This problem will be considered for a dirty system within semiclassical Usadel equations. In the case of  a clean superconductor this problem has been previously considered for a special situation when $\mathbf{Z}$ varies fast within the helix period \cite{MalshArXiv}. $\mathbf{Z}$ will be assumed weak enough, much smaller than $|\Delta|$. In this range it is far below the fields where the so called LOFF \cite{LOFF} state is realized. As it will be shown below, at such small Zeeman fields $|\Delta|$ changes weakly, in comparison with its unperturbed value $|\Delta_0|$.  Therefore, the equation $\Delta=\Delta_0\exp(-i\phi(\mathbf{r}))$ represents a reasonable form of the order parameter. Then, the electric current can be calculated from the Usadel equations. The charge conservation condition $\bm{\nabla}\mathbf{J}=0$ results in an equation that allows to determine the phase $\phi(\mathbf{r})$. This equation has the form of the Laplace equation for $\phi(\mathbf{r})$, where the gradients of $\mathbf{Z}$ play the role of sources. The latter produce a long-range effect on $\phi(\mathbf{r})$, leading to a striking result that either the helix state without the electric current, or an alternative current-carrying state with the constant in space $\phi(\mathbf{r})$ are realized, depending on the shape of a ferromagnetic island. Also, a mixture of these states can be created. Another important characteristic of a superconducting state in such inhomogeneous systems is that the magnetoelectric current is not localized solely inside the ferromagnetic island. It propagates far outside it and decreases according to a power law in the asymptotic range. Such a nonlocality is of particular importance in multiply connected systems, where it results in quantum oscillation effects. An example of a superconducting ring that partly encloses a ferromagnetic island will be discussed briefly. It will be shown that in such a system Little-Parks oscillations, that are produced by a magnetic flux through the ring, are shifted by a constant effective flux determined by a combination of SOI and the Zeeman field. This situation is reminiscent of a constant phase shift in the current-phase characteristics of ferromagnetic spin-orbit coupled Josephson junctions \cite{ISHE,phijunction}.

The article is organized in the following way. In Sec.II the equation for $\phi(\mathbf{r})$ is derived by minimization of the system's thermodynamic potential. In Sec.III the Usadel equations are derived and the magnetoelectric current is calculated. Some examples are considered in Sec.IV, where the helix state and magnetoelectric current are analyzed depending on the shape of a ferromagnetic island and the direction of $\mathbf{Z}$. Sec.V contains a discussion of the results obtained, as well as a brief consideration of quantum oscillations in a superconducting ring. In Appendices the calculations are presented in more detail, in particular, a description is given of  main steps in deriving of the Usadel equations.

\section{Equilibrium phase of the order-parameter}

Within the BCS model we consider a two-dimensional superconductor with the s-wave two-particle attractive interaction
\begin{equation}\label{Hint}
H_{\text{int}}=U\sum_{\mathbf{k,k^{\prime},q}}c^\dag_{\mathbf{k+q},\sigma}c^\dag_{\mathbf{-k},\sigma^{\prime}}c_{\mathbf{-k^{\prime}},\sigma^{\prime}}
c_{\mathbf{k^{\prime}+q},\sigma}\,,
\end{equation}
where $c^\dag_{\mathbf{k},\sigma}$ is the creation operator of a particle with the wave-vector $\mathbf{k}$ and the spin $\sigma$. The summation over the wave-vectors in $H_{\text{int}}$ is restricted to a thin shell around the Fermi level whose width is of the order of the phonon Debye frequency. The Hamiltonian of the system also includes the Zeeman interaction $H_Z=\sum_{\mathbf{k,q}}c^{\dag}_{\mathbf{k+q},\alpha}(\bm{\sigma}_{\alpha\beta}\cdot \mathbf{Z}_{\mathbf{q}})c_{\mathbf{k},\beta}$ of electron spins with the exchange field $\mathbf{Z}(\mathbf{r})$, where $\bm{\sigma}$ is the vector of Pauli matrices. This field is parallel to the superconducting film and can vary in space; $\mathbf{Z}_{\mathbf{q}}$ is the Fourier transform of $\mathbf{Z}(\mathbf{r})$. It can be  either of intrinsic or extrinsic origin. In the latter case it might be induced by the exchange interaction with electrons of a magnetic insulator adjacent to the superconducting film. One more spin-dependent interaction is the spin-orbit coupling which has  the form $H_s=\sum_{\mathbf{k}}c^{\dag}_{\mathbf{k},\alpha}(\bm{\sigma}_{\alpha\beta}\cdot \mathbf{h}_{\mathbf{k}})c_{\mathbf{k},\beta}$. The spin-orbit field $\mathbf{h}_{\mathbf{k}}=-\mathbf{h}_{\mathbf{-k}}$ is assumed to be a linear function of $\mathbf{k}$. This situation takes place if $\mathbf{h}_{\mathbf{k}}$ is represented by the Rashba field \cite{Rashba} $\mathbf{h}_{\mathbf{k}}=\alpha (\mathbf{e}_z\times\mathbf{k})$, or by the linear Dresselhaus \cite{Dresselhaus} field $h_x=\beta k_x, h_y=-\beta k_y$. Both the Zeeman and spin-orbit fields are assumed much smaller than the chemical potential $\mu$. In the case of a uniform in space Zeeman field and Rashba SOI the phase diagram of such a system has been considered in Ref. \onlinecite{Dimitrova}. The superconductor has been shown to be in the Larkin-Ovchinnikov-Fulde-Ferrel phase \cite{LOFF} at $Z\gtrsim |\Delta|$, where $\Delta$ is the superconducting order-parameter. At the same time, at smaller $Z$ the system is in the so called helix phase. In this inhomogeneous state $\Delta(\mathbf{r})$ varies in space as $\Delta_0\exp(-i\mathbf{Q}\mathbf{r})$, where the vector $\mathbf{Q}$  is perpendicular to the Zeeman field. The theory below will be restricted to the helix phase. Therefore, the Zeeman field will be assumed to be smaller than $|\Delta|$.

Within the mean-field approximation the total Hamiltonian $H_0+H_{\text{int}}+H_Z+H_s$ can be written in the form
\begin{equation}\label{H}
H=\sum_{\mathbf{k},\mathbf{k}^{\prime}}\psi^{\dag}_{\mathbf{k}}\mathcal{H}_{\mathbf{kk^{\prime}}}\psi_{\mathbf{k^{\prime}}}\,,
\end{equation}
where  $\psi_{\mathbf{k},\uparrow,1}=c_{\mathbf{k},\uparrow},\, \psi_{\mathbf{k},\downarrow,1}=c_{\mathbf{k},\downarrow},\, \psi_{\mathbf{k},\uparrow,2}=c^{+}_{-\mathbf{k},\downarrow},\, \psi_{\mathbf{k},\downarrow,2}=-c^{+}_{-\mathbf{k},\uparrow}$ with the arrows  and the integers "1" and "2" denoting spin and Nambu variables, respectively. In the spatial representation the  one-particle Hamiltonian takes the form
\begin{equation}\label{Hcal}
\mathcal{H}=\tau_3(\epsilon_{\mathbf{\hat{k}}}-\mu)+\tau_3\mathbf{h}_{\mathbf{\hat{k}}}\bm{\sigma}+\mathbf{Z}(\mathbf{r)}\bm{\sigma}+
\texttt{Re}[\Delta(\mathbf{r})]\tau_1-\texttt{Im}[\Delta(\mathbf{r})]\tau_2\,,
\end{equation}
where $\epsilon_{\mathbf{\hat{k}}}=\hat{k}^2/2m$, $\mathbf{\hat{k}}=-i\partial/\partial\mathbf{r}$ and the Pauli matrices  $\tau_1,\tau_2,\tau_3$ operate in the  Nambu space. In a dirty superconductor $\mathcal{H}$ includes also a random potential associated with impurities.

When $\mathbf{Z}$ is homogeneous in space the helical phase is characterized by the order parameter whose phase depends linearly on $\mathbf{r}$. It is interesting to find out how the phase varies in space when the Zeeman field is nonuniform. A most appropriate trial form of the order parameter in this case is $\Delta(\mathbf{r})=\Delta_0\exp(-2i\phi(\mathbf{r))}$. Here $\Delta_0$ is the real order-parameter in the absence of the Zeeman interaction. In fact, there could be a $Z$- dependent order-parameter. However, when $Z$ is small compared to $|\Delta_0|$ and $h_{\mathbf{k}_F}$, where $k_F$ is the Fermi wave-vector, it produces the major effect on the phase of the order parameter, while its magnitude stays intact, at least up to linear in $Z$ corrections. Such a weak dependence of $|\Delta|$ will be confirmed by calculations in the next section. Further, with such a trial form, we unitary transform Eq.(\ref{Hcal}) as $\mathcal{\tilde{H}}=U\mathcal{H}U^{-1}$, where $U=\exp(i\tau_3\phi(\mathbf{r}))$. The transformed Hamiltonian takes the form
\begin{eqnarray}\label{Htilde}
\mathcal{\tilde{H}}=\tau_3(\epsilon_{\mathbf{\hat{k}}}-\mu)+\tau_3\mathbf{h}_{\mathbf{\hat{k}}}\bm{\sigma}+\mathbf{Z}(\mathbf{r})\bm{\sigma}+
\Delta_0\tau_1 - \nonumber \\
\bm{\nabla}\phi\mathbf{\hat{v}}+\frac{\tau_3}{2m}(\nabla\phi)^2 \,,
\end{eqnarray}
where $\mathbf{\hat{v}}=(\mathbf{\hat{k}}/m)+\bm{\nabla}_{\mathbf{k}}(\mathbf{h}_{\mathbf{k}}\bm{\sigma})$ is the velocity operator. The advantage of the transformed Hamiltonian is that large variations of the order-parameter, that are associated with its phase, are removed. Therefore, with this Hamiltonian one may apply a perturbation expansion with respect to $Z$, which will be assumed small in comparison with $\Delta_0$.

By minimizing the thermodynamic potential with respect to $\phi$ (see  Appendix A) we arrive to the equation
\begin{equation}\label{phi}
\bm{\nabla}\mathbf{J}_v-\frac{en}{m}\nabla^2\phi=0 \,,
\end{equation}
where $\mathbf{J}_v$ is the current given by the thermodynamic average of the velocity operator and $n$ is the electron density. $\mathbf{J}_v$ depends on $Z$ and $\phi$, while the density is fixed by the charge neutrality, at least in the scales much larger than the Fermi wavelength. Note that besides $\mathbf{J}_v$ the total current $\mathbf{J}$ includes also a current associated with the phase gradient, so that Eq.(\ref{phi}) can be written in the form $\bm{\nabla}\mathbf{J}=0$, that is simply a continuity equation for the current.

\section{Magnetoelectric current}

In terms of the thermal Green function the current $\mathbf{J}_{v}$ can be written as
\begin{eqnarray}\label{J2}
\mathbf{J}_v(\mathbf{r})=ek_BT\sum_{\omega_n}\mathrm{Tr}[\frac{i}{2m}(\nabla_{\mathbf{r}^{\prime}}-\nabla_{\mathbf{r}})
G(\mathbf{r},\mathbf{r}^{\prime},\omega_n)|_{\mathbf{r}\rightarrow\mathbf{r}^{\prime}}+\nonumber\\\bm{\nabla}_{\mathbf{k}}(\mathbf{h}_{\mathbf{k}}\bm{\sigma})
G(\mathbf{r},\mathbf{r},\omega_n)] \,,
\end{eqnarray}
where $e$ is the electron charge, the trace is taken over the spin and Nambu variables and $\omega_n=k_BT\pi(2n+1)$, $n=0,\pm 1,\pm 2,..$. In its turn, $G(\mathbf{r},\mathbf{r}^{\prime},\omega_n)$ is determined by the Dyson equation
\begin{equation}\label{G}
(i\omega_n-\mathcal{\tilde{H}}-\Sigma)G(\mathbf{r},\mathbf{r}^{\prime},\omega_n)=
\hat{1}\delta(\mathbf{r}-\mathbf{r}^{\prime})\,,
\end{equation}
where $\hat{1}$ denotes the unit matrix in the spin and Nambu spaces and $\Sigma$ is the self-energy associated with the impurity scattering. After averaging over random positions of short-range impurities in the Born approximation the self-energy takes the form \cite{AGD}
\begin{equation}\label{Sigma}
\Sigma(\mathbf{r},\omega_n)=\frac{1}{2\tau_{\text{sc}}\pi
N_F}\tau_3G(\mathbf{r},\mathbf{r},\omega_n)\tau_3\,,
\end{equation}
where $\tau_{\text{sc}}$ is the elastic scattering time and $N_F$ is the state density at the Fermi level.

Since all relevant energy parameters, such as $|\Delta_0|$, $Z$, $h_{\mathbf{k}_F}$, $1/\tau_{\text{sc}}$ and $v_F\nabla\phi$ are much less than the chemical potential, one may use the semiclassical theory for calculation of $G(\mathbf{r}_1,\mathbf{r}_2,\omega_n)$. In the semiclassical approximation this function varies slowly as a function of the center of gravity $\mathbf{r}=(\mathbf{r}_1+\mathbf{r}_2)/2$. At the same time, as a function of $\mathbf{r}_1-\mathbf{r}_2$ it oscillates fast, within the Fermi wavelength. Therefore, it is convenient to Fourier transform $G$ with respect to $(\mathbf{r}_1-\mathbf{r}_2)$ and retain  intact its dependence on $\mathbf{r}$. Accordingly, let us introduce such a Green function as
\begin{equation}\label{Gk}
G_{\mathbf{k}}(\mathbf{r},\omega_n)=\tau_3\int d^2(\mathbf{r}_1-\mathbf{r}_2)e^{-i\mathbf{k}(\mathbf{r}_1-\mathbf{r}_2)}G(\mathbf{r}_1,\mathbf{r}_2,\omega_n)
\end{equation}
 The semiclassical equation for this function is obtained by subtraction of two Eqs.(\ref{G}) with the operator $(i\omega_n-\mathcal{\tilde{H}}-\Sigma)$  acting on the Green function from the left and from the right, respectively. In addition, coordinate dependent parameters in these equations have to be expanded with respect to small $\mathbf{r}_1-\mathbf{r}_2$. The corresponding procedure is well described in literature \cite{Larkin semiclassical,Rammer}. By using this method  the semiclassical equation for our system is obtained from Eqs. (\ref{Htilde},\ref{G},\ref{Sigma}) in the form
\begin{widetext}
\begin{equation}\label{GZa}
\left[i\hat{\Omega}_n -\mathbf{h}_{\mathbf{k}}\bm{\sigma}+\tau_3\mathbf{v}\bm{\nabla}_{\mathbf{r}}\phi-\tau_3\bm{\sigma}\mathbf{Z}-\Sigma\tau_3 ,G_{\mathbf{k}}\right]=
-\frac{i}{2}\left\{\mathbf{V},\bm{\nabla}_{\mathbf{r}}G_{\mathbf{k}}\right\}+
\frac{i}{2}\left\{\bm{\nabla}_{\mathbf{r}}(\tau_3\bm{\sigma}\mathbf{Z}+\Sigma\tau_3)-\tau_3\nabla^2_{\mathbf{r}}\phi \mathbf{v},\bm{\nabla}_{\mathbf{k}}G_{\mathbf{k}}\right\}\,,
\end{equation}
\end{widetext}
where $\hat{\Omega}_n=\omega_n\tau_3+\Delta_0\tau_2$, $\mathbf{v}=(\mathbf{k}/m)+\bm{\nabla}_{\mathbf{k}}(\mathbf{h}_{\mathbf{k}}\bm{\sigma})$, and $\mathbf{V}=\mathbf{v}-(\tau_3/m)\bm{\nabla}_{\mathbf{r}}\phi$. For brevity some of the  arguments have been omitted in $G_{\mathbf{k}}(\mathbf{r},\omega_n),\mathbf{Z}(\mathbf{r})$, $\phi({\mathbf{r}})$ and $\Sigma(\mathbf{r})$. Usually, the semiclassical equations like Eq.(\ref{GZa}) are integrated over $\xi\equiv\epsilon_k-\mu$ to get the Eilenberger equation \cite{Eilenberger,Larkin semiclassical,Rammer} for an integrated over $\xi$ Green's function with a fixed direction of the Fermi momentum. By this way an analysis of semiclassical equations may be tremendously simplified.  It can not be done in our case, because this method requires a fixation at the Fermi surface of all wave-vector dependent parameters in the equation. In our case, however,  it is important to resolve two Fermi surfaces corresponding to spin-split electron bands where electrons at fixed $\mathbf{k}$ have slightly different velocities $\mathbf{v}_F\pm \bm{\nabla}_{\mathbf{k}}h_{\mathbf{k}}$. This problem with semiclassical equations is common for many spin-orbit effects where the coupling of spin and charge degrees of freedom is important. A method for derivation of Eilenberger equations that is based on the non-Abelian gauge theory has been proposed for normal \cite{Gorini} and superconducting \cite{Bergeret} spin-orbit coupled systems. In practice, such a theory cannot avoid an expansion over the small nonclassical parameter $h_{\mathbf{k}}/\mu$. In our case, for calculation of the magnetoelectric current it is sufficient to take into account the spin-splitting effects up to the first order with respect to this parameter. With this accuracy the linearized with respect to  small $Z\ll min[\Delta_0,h_{\mathbf{k}_F}]$ Usadel equations can be directly obtained from Eq.(\ref{GZa}), without transforming the latter to the Eilenberger equation.

In the regime of strong disorder, which will be considered below, the elastic scattering time $\tau_{\text{sc}}$ is short, so that $\tau_{\text{sc}} E \ll 1$, where $E$ is any of the energy parameters $Z, \Delta_0, h_{\mathbf{k}}$ and $ v_F/L$, with $L$ denoting the scale of spatial variations of the Zeeman field. Strong elastic scattering leads to a fast randomization of the particle's direction of motion. As a result, $G_{\mathbf{k}}(\mathbf{r},\omega_n)$ is almost isotropic in $\mathbf{k}$. At the same time, higher angular harmonics contain higher powers of the small parameter $\tau_{\text{sc}}E$. By treating these harmonics perturbatively it is possible to derive a closed diffusion equation for the  function $g(\mathbf{r},\omega_n)$ defined as $g(\mathbf{r},\omega_n)=(\pi N_F)^{-1}\sum_{\mathbf{k}}G_{\mathbf{k}}(\mathbf{r},\omega_n)$. The procedure of derivation of this, so called Usadel equation from the Eilenberger equation  is well described in literature \cite{Larkin semiclassical,Rammer}. For the considered here system a set of coupled Usadel equations will be derived directly from Eq.(\ref{GZa}), as presented in more detail in Appendix B. For simplicity, $\mathbf{h}_{\mathbf{k}}$ will be taken in the form of the Rashba interaction. It is convenient to represent $g(\mathbf{r},\omega_n)$ as  $g=g_s+g_{\parallel}\bm{\kappa}\bm{\sigma}+
g_{\perp}\bm{\nu}\bm{\sigma}+g_{z}\sigma_z$ where $\bm{\kappa}=\mathbf{q}/q$ and $\bm{\nu}=\hat{\mathbf{z}}\times \bm{\kappa}$, with $\hat{\mathbf{z}}$ denoting the unit vector in the $z$-direction. In  the Nambu space a basis will be chosen such that the operator $\hat{\Omega}_n\equiv\tau_3\omega_n +\tau_2\Delta_0$ is diagonal. In such a basis the linearized with respect to Fourier components $\mathbf{Z}_{\mathbf{q}}$ and $\phi_{\mathbf{q}}$ Usadel equations take the form
\begin{eqnarray}\label{Usadel}
&&(2\Omega_n+Dq^2)g_{s}-2i\tau_3\alpha q h^2_{k_F}\tau_{\text{sc}}^2g_{\perp} =-2i\tau_1Dq^2\frac{\Delta_0}{\Omega_n}\phi_{\mathbf{q}}\,,\nonumber \\
&&(2\Omega_n+Dq^2+\Gamma_s)g_{\parallel}+4i\alpha m Dqg_{z}=-2\tau_2\frac{\Delta_0}{\Omega_n}\mathbf{Z}_{\mathbf{q}}\bm{\kappa}\,,\nonumber \\
&&(2\Omega_n+Dq^2+\Gamma_s)g_{\perp}-2i\tau_3\alpha q h^2_{\mathbf{k}_F}\tau_{\text{sc}}^2g_{s}=\nonumber\\
&&4\tau_2\alpha q h^2_{\mathbf{k}_F}\tau_{\text{sc}}^2\frac{\Delta_0}{\Omega_n}\phi_{\mathbf{q}}-2\tau_2\frac{\Delta_0}{\Omega_n}
\mathbf{Z}_{\mathbf{q}}\bm{\nu}\,,\nonumber \\
&&(2\Omega_n+Dq^2-2\Gamma_s)g_{z}-4i\alpha m Dqg_{\parallel}=0\,,
\end{eqnarray}
where $\Gamma_s=2h_{k_F}^2\tau_{\text{sc}}$ is the D'yakonov-Perel' spin-relaxation rate \cite{DP}, $D=v_F^2\tau_{\text{sc}}/2$ is the diffusion constant and $\Omega_n=\sqrt{\omega_n^2+\Delta_0^2}$.

Eqs. (\ref{Usadel}) are diffusion equations for the singlet and triplet pairing functions. The triplet functions $g_{\parallel}$ and $g_{z}$ couple to each other due to spin precession in the spin-orbit field. All three triplet functions are nonzero due to the Zeeman interaction in the right-hand side (r.h.s.) of  Eqs.(\ref{Usadel}). This spin polarization is then communicates via SOI to the electric current Eq.(\ref{jnu}). The $g_{\perp}$-triplet  contributes to the first Eq.(\ref{Usadel}) for the singlet function through the singlet-triplet coupling term (the second term in the first equation). At the same time, the corresponding conjugate term couples $g_{\perp}$ to $g_s$ in the third equation. These couplings, as well as the coupling of $g_{\perp}$ to $\phi_{\mathbf{q}}$ in the r.h.s of the third equation are small. They contain the small nonclassical parameter $\alpha/v_F\sim h_{\mathbf{k}}/\mu$ and they are the only nonclassical terms  in Eq.(\ref{Usadel}). This singlet-triplet, or spin-charge, mixing is responsible for a number of spin-related phenomena in superconducting and normal systems. The coupling of  $g_s$ to $g_{\perp}$ in the first equation gives rise to the helix rotation of the order-parameter, as it will be clear below. The conjugate term in the third equation is associated with the spin-Hall effect \cite{SHE} in Josephson junctions, while the coupling to the phase $\phi_{\mathbf{q}}$ in this equation allows to induce the spin polarization by a condensate current.\cite{Yip,EdelsteinPRL} In normal metals there are similar terms that couple spin and charge densities in diffusion equations. \cite{Mishchenko}  It is noteworthy that at $\mathbf{Z}=0$ and $\phi_{\mathbf{q}}=0$ Eqs.(\ref{Usadel}) for the triplet pairing function coincide with diffusion equations for the spin-density in normal metals \cite{Mishchenko}. There  $-i\Omega_n$ is substituted for the real-time frequency and $g_s$ for $eVN_F$, where $V$ is the potential of the external electric field. For the considered here problem one may neglect nonclassical terms in the third equation (\ref{Usadel}), because they are small by the parameter $(\alpha/v_F)^2$ ($\rho_s$ and $\phi_{\mathbf{q}}$  are small as $\alpha/v_F$).

The electric current can be expressed from Eq.(\ref{J2}) in terms of the mixed Fourier transformed Green function Eq.(\ref{Gk}) as
\begin{equation}\label{j4}
\mathbf{J}_v(\mathbf{r})=e\frac{k_BT}{2}\sum_{\mathbf{k},\omega_n} \mathrm{Tr}[\tilde{\tau}_3\mathbf{v}G_{\mathbf{k}}(\mathbf{r},\omega_n)]\,,
\end{equation}
where $\tilde{\tau}_3=(\omega_n/\Omega_n)\tau_3-(\Delta_0/\Omega_n)\tau_2$ is the Pauli matrix $\tau_3$ rotated to the diagonal with respect to $\hat{\Omega}_n $ basis and the trace is taken over Nambu and spin variables. The major contribution to the sum over $k$ in this expression is given by the region close to the Fermi wave-vector $k_F$. However, due to the term $\bm{\nabla}\phi \hat{\mathbf{v}}$ in Hamiltonian (\ref{Htilde}) a region of the $k$-space far from the Fermi surface also contributes to the current. In this region $G_{\mathbf{k}}(\mathbf{r},\omega_n)$ coincides with the Green function  of the normal metal. At the same time, $\bm{\nabla}\phi \hat{\mathbf{v}}$ formally looks as an interaction with the electromagnetic vector potential $(e/c)\mathbf{A}=\bm{\nabla}\phi$. As well known, in a normal metal Eq.(\ref{j4}) plus the diamagnetic current must be zero in the thermal equilibrium. Therefore, the second "diamagnetic" term in Eq.(\ref{phi}) is canceled due to the contribution in Eq.(\ref{j4}) of large $|k-k_F|$. Hence, only the range close to $k_F$ will be taken into account in the sum over $\mathbf{k}$ in Eq.(\ref{j4}) and the second term in Eq.(\ref{phi}) will be omitted. The details of such a procedure can be found in Ref.\onlinecite{Kopnin}. So defined $\mathbf{J}_v(\mathbf{r})$ coincides with the total current $\mathbf{J}(\mathbf{r})$. Therefore, in $\mathbf{J}_v(\mathbf{r})$ the subscript $v$ will be omitted below .

Within the diffusion approximation $G_{\mathbf{k}}(\mathbf{r},\omega_n)$ in Eq.(\ref{j4}) can be expressed through $g(\mathbf{r},\omega_n)$ and the sum over  $\mathbf{k}$ may be explicitly calculated, as shown in Appendix B. It is convenient to decompose the Fourier components of the current into the longitudinal $\mathbf{J}_{\mathbf{q}\parallel}=\bm{\kappa}(\mathbf{J}_{\mathbf{q}}\bm{\kappa})$ and transverse  $\mathbf{J}_{\mathbf{q}\perp}=\bm{\nu}(\mathbf{J}_{\mathbf{q}}\bm{\nu})$ parts. By this way  the current may be expressed in the form
\begin{eqnarray}\label{jnu}
\mathbf{J}_{\mathbf{q}\perp}&=&
-\bm{\nu}\tau_{\mathrm{sc}}C\sum_{\omega_n}\frac{\Delta_0}{\Omega_n}\mathrm{Tr}[\tau_2(\alpha\Gamma_sg_{\parallel}+2i\alpha^2 qmDg_{z})]\,, \nonumber\\
\mathbf{J}_{\mathbf{q}\parallel}&=&
\bm{\kappa}C\sum_{\omega_n}\frac{\Delta_0}{\Omega_n}\mathrm{Tr}[\tau_2(\tau_{\mathrm{sc}}\alpha\Gamma_sg_{\perp}+iqD\tau_3 g_{s}- \nonumber \\
&&2iD\tau_2\frac{\Delta_0}{\Omega_n}q\phi_{\mathbf{q}})]\,,
\end{eqnarray}
where $C=e\pi N_Fk_BT$. The longitudinal and transverse currents can easy be obtained by calculating the  $s,\parallel,\perp$ and $z$-components of $g$ from Usadel equations (\ref{Usadel}). Let us first consider the longitudinal current given by the second line of Eq.(\ref{jnu}).  By taking into account that for Rashba SOI $\alpha\mathbf{Z}_{\mathbf{q}}\bm{\nu}=\bm{\kappa}\bm{\nabla}_{\mathbf{k}}(\mathbf{h}_{\mathbf{k}}\mathbf{Z})$ this current can be written as
\begin{widetext}
\begin{equation}\label{jpar}
\mathbf{J}_{\mathbf{q}\parallel}=-8C\bm{\kappa}(\bm{\kappa}\bm{\nabla}_{\mathbf{k}})(\mathbf{h}_{\mathbf{k}}\mathbf{Z}_{\mathbf{q}})\tau_{\mathrm{sc}}
\sum_{\omega_n}\frac{\Delta_0^2\Gamma_s}{\Omega_n(2\Omega_n+Dq^2+\Gamma_s)(2\Omega_n+Dq^2)}-8Ci\mathbf{q}D\phi_{\mathbf{q}}
\sum_{\omega_n}\frac{\Delta_0^2}{\Omega_n(2\Omega_n+Dq^2)}\,.
\end{equation}
\end{widetext}
This longitudinal current  must be zero. It is required by the charge conservation law  Eq.(\ref{phi}). From the equation $\mathbf{q}\mathbf{J}_{\mathbf{q}\parallel}=0$ we obtain the equation for determining the phase $\phi_{\mathbf{q}}$. While the longitudinal current turns to zero, the transverse one is finite. This current contributes to the magnetoelectric effect. For the transverse current Eqs.(\ref{jnu}) and (\ref{Usadel}) give
\begin{widetext}
\begin{equation}\label{jperp}
\mathbf{J}_{\mathbf{q}\perp}=-8C\bm{\nu}(\bm{\nu}\bm{\nabla}_{\mathbf{k}})(\mathbf{h}_{\mathbf{k}}\mathbf{Z}_{\mathbf{q}})\tau_{\mathrm{sc}}
\sum_{\omega_n}\frac{\Delta_0^2}{\Omega_n^2}
\frac{\Gamma_{s}(2\Omega_n-Dq^2 +2\Gamma_{s})}{[(2\Omega_n+Dq^2+\Gamma_{s})(2\Omega_n+Dq^2+2\Gamma_{s})-4\Gamma_{s}Dq^2]}\,.
\end{equation}
\end{widetext}

Before concluding this section let us check, if the linear in $Z$ correction to the order-parameter $\Delta_0$ is  absent, as was assumed above.  With isotropic and spin-independent electron-electron attractive interaction $(\ref{Hint})$ this correction is given by the sum over $\omega_n$ of the anomalous part of $g_s(\mathbf{r},\omega_n)$. According to Eq.(\ref{Usadel}), $g_s(\mathbf{r},\omega_n)$ is proportional to $\tau_1$ and, hence, is nondiagonal in the Nambu space. Therefore, it might contribute to the order-parameter. However, it is easy to check that by taking into account Eq.(\ref{jpar}) and $\mathbf{J}_{\mathbf{q}\parallel}=0$, a direct calculation of $\sum_n g_s(\mathbf{r},\omega_n)$  from Eq.(\ref{Usadel}) will give a zero result.  Hence, the assumption that $\Delta_0$ in Eq.(\ref{Htilde}) is fixed, up to the linear with respect to $Z$ terms, was correct.

\section{Helix phase vs magnetoelectric effect}

It is easy to see from Eq.(\ref{jpar}) that the equation $\mathbf{q}\mathbf{J}_{\mathbf{q}\parallel}=0$ has the form of the Laplace equation for the phase $\phi$. At the same time, the $Z$-dependent first term in Eq.(\ref{jpar}) plays the role of the external source. By drawing an analogy with electrostatics one may conclude that the phase $\phi(\mathbf{r})$ may depend on spatial variations of $\mathbf{Z}(\mathbf{r}^{\prime})$ at $\mathbf{r}^{\prime}$ quite distant from  $\mathbf{r}$. Therefore, without analyzing variations of  $\mathbf{Z}(\mathbf{r})$ in the whole sample, it is impossible to say unambiguously if one will observe the helix phase, or the magnetoelectric supercurrent, or both. In order to illustrate this point let us consider a superconducting film where $\mathbf{Z}(\mathbf{r})$ is uniform inside a region of a rectangular shape. The Zeeman field can have two orientations, as shown in Fig.(1). The lengths of rectangular sides will be assumed much larger than the spin-orbit length $L_s=2\sqrt{D/\Gamma_s}=(m\alpha)^{-1}$ and the coherence length $L_c=\sqrt{D/\Delta_0}$. Also, we are going to calculate $\phi(\mathbf{r})$ and $\mathbf{J}_{\mathbf{q}}$ far enough from the rectangular boundary, with respect to these two lengths.  Therefore, one may set $q=0$ in denominators of Eqs.(\ref{jpar}-\ref{jperp}). After this simplification, Eq.(\ref{jpar}) results in the continuity equation
\begin{equation}\label{Laplace}
\mathbf{\rho}_Z+\nabla^2\phi=0\,,
\end{equation}
where the effective "charge" density $\mathbf{\rho}_Z$ is given by
\begin{equation}\label{rhoZ}
\rho_Z=-\frac{(\bm{\nabla}_{\mathbf{r}}\bm{\nabla}_{\mathbf{k}})\left(\mathbf{h}_{\mathbf{k}}\mathbf{Z}(\mathbf{r})\right)\tau_{\mathrm{sc}}}
{D\sum_{\omega_n}\Delta_0^2\Omega_n^{-2}}
\sum_{\omega_n}\frac{\Delta_0^2}{\Omega_n^2}
\frac{\Gamma_{s}}{(2\Omega_n+\Gamma_{s})}\,.
\end{equation}
The current, in turn, is determined by Eq.(\ref{jperp}). This expression, however, is not very convenient, because in the coordinate representation it has a nonlocal form. A more transparent equation can be obtained by adding to Eq.  (\ref{jperp}) a zero longitudinal current Eq.(\ref{jpar}). At small $q$ the sum of two $Z$-dependent terms in these equations has a simple local form that will be denoted as $\mathbf{J}_0$. In the coordinate representation the latter is given by
\begin{equation}\label{J0}
\mathbf{J}_0(\mathbf{r})=-4C\bm{\nabla}_{\mathbf{k}}(\mathbf{h}_{\mathbf{k}}\mathbf{Z}(\mathbf{r}))\tau_{\mathrm{sc}}
\sum_{\omega_n^2}\frac{\Delta_0^2\Gamma_s}{\Omega_n^2(2\Omega_n+\Gamma_s)}\,.
\end{equation}
Accordingly, the equation for the current takes the form
\begin{equation}\label{Jfin}
\mathbf{J}(\mathbf{r})=\mathbf{J}_0(\mathbf{r})-\frac{en_s}{m}\bm{\nabla}\phi(\mathbf{r})\,,
\end{equation}
where $n_s=4(m/e)CD\sum_{\omega_n}(\Delta_0^2/\Omega_n^2)$ is a 2D density of superconducting electrons.\cite{Kopnin} Eq.(\ref{Jfin}) has a clear physical meaning. The first term represents the magnetoelectric current directly induced by the magnetic field. This term has been calculated for a clean superconductor in Refs.\onlinecite{Edelstein,Yip}. The second term is the current produced by the order-parameter phase gradient. It is also useful to have an integral form of Eq.(\ref{Jfin}). By applying a $curl$ operation to this equation and integrating it over an area enclosed by the contour  $\mathcal{C}$ we arrive to the equation
\begin{equation}\label{Jcurl}
\oint_{\mathcal{C}} d\mathbf{r}\mathbf{J}(\mathbf{r})=\int d^2r B\,,
\end{equation}
where $B=(\bm{\nabla}\times\mathbf{J}_0)_z$. It is important that the expression in the r.h.s. is not zero only if $\mathbf{Z}$ and, hence, $\mathbf{J}_0)_z$ varies in space. Hence, a finite supercurrent may be induced only by a varying in space Zeeman field. For calculation of the current it is convenient to transform Eq.(\ref{Jfin}) in the following way. Let us introduce the function $\chi(\mathbf{r})$, so that $\mathbf{J}(\mathbf{r})=\bm{\nabla}\times\mathbf{e}_z\chi(\mathbf{r})$, where $\mathbf{e}_z$ is the unit vector parallel to the $z$-axis. By calculating $\bm{\nabla}\times \mathbf{J}$ in Eq.(\ref{Jfin}) we obtain the equation
\begin{equation}\label{Laplace2}
B+\nabla^2\chi=0\,,
\end{equation}
This equation is auxiliary to Eq.(\ref{Laplace}). Both equations will be applied for analysis of the examples shown in Fig.1. It should be noted that $\mathbf{J}$ induces the magnetic field which, in turn, induces the current proportional to the vector-potential of this field. One must add this current to Eq.(\ref{Jfin}). Then, the total current may be inserted into the Maxwell equation. From this equation the screening magnetic field can be determined. In the case of  thin films its effect can be ignored at distances much less than $L_B^2/d$, where $L_B$ is the magnetic field penetration length and $d$ is the film thickness. \cite{Pearl} The latter is small for the considered 2D system, while $L_B$ is much larger than $d$. Therefore, the screening effects are important only at very large distances.
\begin{figure}[tp]
\includegraphics[width=8.8cm]{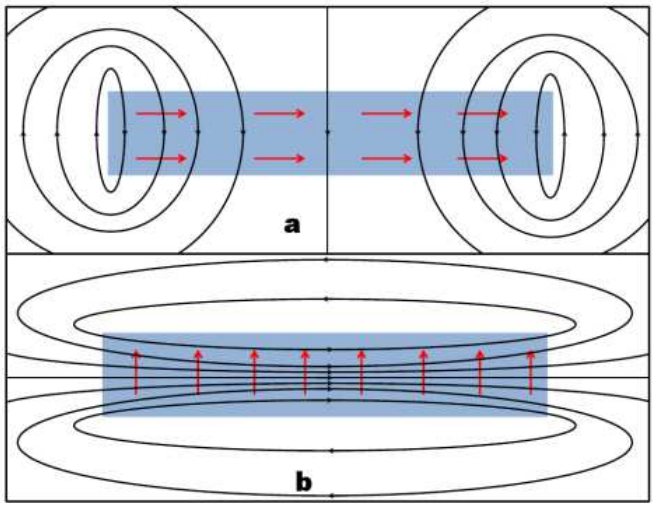}
\caption{(Color online) A rectangular bar depicts an island inside a 2D superconductor where the Zeeman coupling is not zero (a direction of the Zeeman field is shown by arrows). Such an inhomogeneous coupling can be produced by a magnetic insulator film on the top of a superconductor film. a) The supercurrent (shown by ovals) is mainly induced near vertical edges of the island. The helix-like variations of the order-parameter take place in the vertical direction nearby its middle (see text).  b) The supercurrent is strong inside the island, closer to its middle, while the order-parameter phase varies strongly near its vertical edges. In the middle the phase varies slowly around zero.} \label{fig1}
\end{figure}

Let us first consider the ferromagnetic island in Fig.1a. Since the Zeeman field turns to zero stepwise on the rectangle's sides, both functions $\rho_Z$ and $B$ are concentrated near these sides and can be represented by delta-functions. In the case of  Rashba coupling it is seen from Eq.(\ref{rhoZ}) that for $\mathbf{Z}$ oriented in the horizontal direction (x-axis) $\rho_Z$ is not zero at two horizontal sides, where it has opposite signs. Therefore, according to Eq.(\ref{Laplace}) spatial variations of $\phi$  have the same form as that of the electric potential around a two-dimensional capacitor. For the island that is strongly elongated along the x-axis one may neglect the effect of edges in the points which are distant enough from them. In these points Eq.(\ref{Laplace}) has a simple solution $\phi=Q y$ inside the island and $\phi=\mathrm{sign}(y) Qw/2$ just outside it, where  $w$ is the vertical size of the rectangle.  In the case of  Rashba SOI $Q$ is given by
\begin{equation}\label{Q}
Q=-\frac{\alpha Z \tau_{\mathrm{sc}}}
{D\sum_{\omega_n}\Delta_0^2\Omega_n^{-2}}
\sum_{\omega_n}\frac{\Delta_0^2}{\Omega_n^2}
\frac{\Gamma_{s}}{(2\Omega_n+\Gamma_{s})}\,.
\end{equation}
Unlike $\rho_Z$, the source $B$ in Eq.(\ref{Laplace2}) is distributed on the vertical edges of the island. The current, in turn, is given by the vector product of $\mathbf{e}_z$ with the gradient of $\chi$. Therefore, it is mostly concentrated near these edges, as shown in Fig. 1a. Since, $\chi \propto \ln r$, where $r$ is a distance from the point-like source, the current decreases as $1/r$ at a distance from it. Such a behavior can also be seen from Eq.(\ref{Jcurl}). Hence, at  $w/l \ll 1$, where $l$ is the horizontal size of the island, the current is vanishing away from the vertical edges. This is just the situation where the helix state \cite{Edelstein,Samokhin,Dimitrova} is realized. Indeed, as expected for such a state, the magnetoelectric current is absent and the order parameter varies in space as $\Delta_0\exp(-2i\phi)=\Delta_0\exp(-2iQy)$. At strong SOI when $\Gamma_s \gg \Delta_0$ Eq.(\ref{Q}) gives $Q=2\alpha Z/v_F^2$ that corresponds to the helix wave-vector at the large spin-orbit coupling \cite{Samokhin,Dimitrova}. However, this phase is realized only in a special geometry, far enough from the edges of a strongly elongated island, where the Zeeman field is directed parallel to its long side (for Rashba spin-orbit coupling). Otherwise, the current is not zero and the phase varies in a manner that is quite different from a simple linear dependence.

In Fig.1b the Zeeman field is parallel to the $y$-axis. In this case $\rho_Z$ is distributed at the vertical sides, while $B$ is concentrated at horizontal ones. At $l \gg w$, far enough from the short edges, the function $\chi(y)$ can be easy obtained from Eq.(\ref{Laplace2}) and the current $J^x$ takes the form $J^x=-\partial\chi/\partial y=J^x_{0}$, where $\mathbf{J}_0$ is given by Eq.(\ref{J0}). This result also  follows directly from Eq.(\ref{Jfin}). Indeed, $\bm{\nabla}\phi$ decreases as $1/r$ at the distance from the vertical edges and can be neglected in Eq.(\ref{Jfin}). This situation in some sense is opposite to the helix state realized for $l \gg w$ in Fig. 1a, because now we do not have considerable spatial variations of the order-parameter phase, while the  magnetoelectric current is finite. In order to get some idea about the magnitude of the magnetoelectric current, let us assume that the spin-orbit coupling is strong, so that $\Gamma_s \gg \Delta_0$ in Eq.(\ref{J0}). At the low temperature $k_B T \ll \Delta_0$ we then obtain from Eq.(\ref{J0}) $\mathbf{J}_0 =-eL^{-1}_s (Z\tau_{\mathrm{sc}})\Delta_0$. Taking $Z\tau_{\mathrm{sc}}=0.1$ and $\Delta_0=$1 meV  the current in a 2D strip  of the width $w$ in Fig. 1b can be evaluated as $wJ_0\approx (w/L_s)$ 4$\cdot$10$^{-9}$ Amp. Depending on the material, the spin-orbit length $L_s$ may vary over a broad range, from 10$^3${\AA} in semiconductor quantum wells, down to nanometers in metals and some conducting oxide interfaces. Therefore, the parameter $w/L_{s}$ may be quite large even in nanosized samples. One should not forget, however, that the theory is valid only for dirty samples where $L_{s} \gg v_F\tau_{\mathrm{sc}}$. It should be noted that in small samples the diffusion effects, which have been ignored in this section, must be taken into account. These effects are represented by $Dq^2$ terms in  Eqs.(\ref{jperp}) and (\ref{jpar}).

\section{Discussion}

The considered above simple examples give an idea of how the  supercurrent can be induced by a static Zeeman field. They demonstrate on a qualitative level in what kind of geometries the magnetoelectric effect can be observed. In the case of an arbitrary function $\mathbf{Z}(\mathbf{r})$ a spatial distribution of the current and the phase of the order parameter can be calculated from Eq.(\ref{jperp}) and the continuity equation $\mathbf{J}_{\mathbf{q}\parallel}=0$, where $\mathbf{J}_{\mathbf{q}\parallel}$ is given by Eq.(\ref{jpar}). The most striking conclusion, that follows from the above analysis, is that inside a macroscopic region, where the Zeeman field is constant in space, the order-parameter phase and supercurrent depend on spatial variations of $\mathbf{Z}(\mathbf{r})$ at a large distance from this region. Therefore, a realization of the helix state that was predicted in Refs.\onlinecite{Edelstein,Samokhin,Dimitrova} for systems having uniform $\mathbf{Z}(\mathbf{r})$ is not guaranteed by a macroscopically local homogeneity of $\mathbf{Z}(\mathbf{r})$. On the other hand, our analysis was based on the assumption that the Zeeman field vanishes at infinity, while the superconductor is boundless. In other words, we have dealt with a ferromagnetic island embedded into superconductor. An alternative set up, however, could be considered. For example, in Fig.1 a ferromagnetic rectangle's boundary might also be a superconductor's boundary. In this case $\rho_Z=B=0$  in Eqs.(\ref{Laplace}) and  (\ref{Laplace2}). Instead, the boundary condition is imposed that the normal projection of $\mathbf{J}$ vanishes on the boundary. For a uniform $\mathbf{Z}$ the corresponding solution of Eqs.(\ref{Laplace}) and  (\ref{Laplace2}) is evident. It is $\chi=0$, $\mathbf{J}=0$ and $\mathbf{J}_0(\mathbf{r})=(en_s/m)\bm{\nabla}\phi(\mathbf{r})$, as it follows from Eq.(\ref{Jfin}). Since $\mathbf{J}_0(\mathbf{r})$ is constant in space, $\bm{\nabla}\phi(\mathbf{r})$ is also constant and Eq.(\ref{Laplace}) is satisfied at $\rho_Z=0$. This means that $\phi$ is a linear function of $\mathbf{r}$. Hence, the helix state with the zero current is realized in such a bounded sample.

Looking at the vortex-like structure of the superconducting current induced around ferromagnetic island edges one might wonder, if quantum oscillation phenomena can be observed in multiply connected systems. Let us consider a simple example of a thin superconducting 2D ring enclosing one of the vertical rectangle's sides in Fig.1a. If the side length $w\ll R$, where $R$ is the ring radius, the current lines in Fig.1(a) have almost a circular form. The current is given by Eqs.(\ref{Laplace}) and (\ref{Jfin}). However, besides a phase produced by the source $\rho_Z$, in such a doubly connected geometry the phase $N\theta/2$ also must be added to $\phi$, where $N$ is an integer and $\theta$ is the polar angle. This additional phase is chosen such that the order-parameter $\Delta \sim \exp(-2i\phi)$ is a periodic function on the circle. This phase contributes $-(en_s/2m)(N/R)$ to the current density (\ref{Jfin}). Hence, (\ref{Jfin}) gives $J=J_Z-(en_s/2m)(N/R)$, where $J_Z$ can be calculated from Eq.(\ref{Jcurl}). By expressing the integral of $B$ in terms of an effective flux, as $(e^2n_s/mc)\Phi_{\mathrm{eff}}$ one obtains $J=(e^2n_s/2\pi mcR)(N\Phi_0-\Phi_{\mathrm{eff}})$, where $\Phi_0=hc/2e$ is the flux quantum. $N$, in turn,  must be determined by a minimization of the condensate's kinetic energy. This situation resembles the Little-Parks effect produced by a magnetic flux piercing a hollow thin walled cylinder,\cite{Little} where periodic oscillations of the current and the superconducting transition temperature have been observed as a function of the magnetic flux. Indeed, the additional magnetic flux $\Phi$ to our ring will result in $J \sim (N\Phi_0-\Phi_{\mathrm{eff}}-\Phi)$. Therefore, the Little-Parks oscillations will be shifted by $\Phi_{\mathrm{eff}}$. $\Phi_{\mathrm{eff}}$, in turn, can be varied by changing the magnetization direction. From the practical point of view there are interesting opportunities in combining magnetic and superconducting circuits, when magnetic islands are incorporated into flux qubit or SQID systems.

It is interesting to evaluate a typical ratio $\Phi_{\mathrm{eff}}/\Phi_0$. For a geometry in Fig.1b, where a superconducting loop of an arbitrary shape encloses one of the rectangle's long sides, at $\Gamma_s \gg \Delta_0$ it is easy to obtain $\Phi_{\mathrm{eff}}/\Phi_0=(Z/\pi \mu)(l/L_s)$. Since our theory is restricted to $Z\ll \Delta_0$, for typical superconductors we have a very small ratio $(Z/\pi \mu) \lesssim 10^{-4}$. Therefore, only a very strong spin-orbit coupling, with the spin-orbit length in the range of nanometers can provide  $\Phi_{\mathrm{eff}}/\Phi_0 \sim 1$ for large enough ferromagnetic islands, whose size $l$ fells  into the micrometer range. A serious restriction on  $\Phi_{\mathrm{eff}}$ is imposed by small values of $Z$ that must be less than $\sim \Delta_0$, because otherwise the system transfers into the LOFF state. In principle, larger $Z$ can be reached in very thin ferromagnetic films in a proximity contact with a massive superconductor. Due to such a contact a finite pairing amplitude takes place inside the film, even at large $Z$. However, one cannot guarantee the large enough magnetoelectric effect in such a system.

Besides the quantum effects that can be observed in multiply connected samples, there are more direct ways to detect the spontaneous currents produced by the Zeeman interaction. Such currents induce magnetic fields. These weak fields might be detected by SQIDs placed just on top of a hybrid planar system. Stronger fields can be produced in multilayer structures. From the above analysis one may expect that the z-components of these fields must be strongest near the island edges that are perpendicular to the Zeeman field.

It should be noted that the static magnetoelectric effect can be realized in  superconducting systems that are quite distinct from the system considered above. The important role of Zeeman field spatial variations to induce the magnetoelectric current has been demonstrated in a one-dimensional topological superconducting junction. \cite{topophijun} The junction is represented by a combination M-S-\~{M}, where M and \~{M} contain Zeeman fields that are rotated with respect to each other. The  situation considered in these papers cannot in principle be projected onto the discussed here problem, even if one choose the parameters such that the system will stay in the non-topological phase. In our case the magnetoelectric current is linear in the Zeeman field and is proportional to $\bm{\nabla}_{\mathbf{k}}(\mathbf{h}_{\mathbf{k}}\mathbf{Z}(\mathbf{r}))$. In \cite{topophijun} the varying in space part of $\mathbf{Z}$ is perpendicular to $\mathbf{h}_{\mathbf{k}}$, while a direction of the latter  does not depend on $\mathbf{k}$. Therefore,   the above formula predicts that such a rotating Zeeman field in the 1D system cannot result in the electric current. This means that the considered in Ref.\onlinecite{topophijun} magnetoelectric effect  belongs to a different class.

%%%%%%%%%%%%%%%%%%%%%%%%%%%%%%%

%%%%%%%%%%%%%%%%%%%%%%%%%%%%%%%%%%%%%%%%%%%%%%%%%%%%%%%%
\appendix

\section{Equation for the phase of the order parameter}

A variation with respect to $\phi(\mathbf{r})$ of the thermodynamic potential $\Omega=-k_BT\ln S$, where $S$ is the grand statistical sum, can be calculated straight by varying Hamiltonian (\ref{Htilde}).\cite{AGD} Accordingly, the corresponding variation can be written in the form
\begin{equation}\label{deltaomega}
\delta\Omega=\int d^2r \left(-\frac{1}{e}\mathbf{J}_{v}\bm{\nabla}\delta\phi(\mathbf{r})+\frac{n}{m}\bm{\nabla}\phi(\mathbf{r})\bm{\nabla}\delta\phi(\mathbf{r})\right)\,,
\end{equation}
where $\delta\phi(\mathbf{r})$ is the phase variation, $\mathbf{J}_v$ is the current given by Eq.(\ref{J2}) and $n=k_BT\sum_{\omega_n}\mathrm{Tr}[\tau_3G(r,r,\omega_n)]$ is the particle density expressed in terms of the thermal Green function $G$. In the thermal equilibrium $\delta\Omega=0$. Since the Zeeman interaction and, hence, $\mathbf{J}_{v}$ and $\phi(\mathbf{r})$ vanish at infinity, one can integrate Eq.(\ref{deltaomega}) by parts, that results in Eq.(\ref{phi}).

\section{Usadel equations}

In this Appendix we will derive Usadel equations from semiclassical Eq.(\ref{GZa}). These equations are valid in a dirty superconductor where the elastic scattering rate $1/\tau_{\text{scat}}$ is much larger than $\Delta_0$, $h_{\mathbf{k}}$ and $\mathbf{vq}$. Eq.(\ref{GZa}) can be preliminary simplified by linearizing it with respect to small $Z$. Accordingly, the Green function $G_{\mathbf{k}}(\mathbf{r},\omega_n)$ that is given by Eq.(\ref{Gk}) may be represented in terms of the unperturbed function $G^0$ and the linear in $\mathbf{Z}$ perturbation $G_Z$ as
\begin{equation}\label{G01}
G_{\mathbf{k}}(\mathbf{r},\omega_n)=\tau_3G^0_{\mathbf{k}}(\omega_n)+G_{\mathbf{k}Z}(\mathbf{r},\omega_n)\,,
\end{equation}
where $G^0$ can be calculated from Eq.(\ref{G}) at $Z=0$. By Fourier transforming Eq.(\ref{G}) with respect to $\mathbf{r}-\mathbf{r}^{\prime}$ we obtain
\begin{widetext}
\begin{equation}\label{G0}
\tau_3G^0_{\mathbf{k}}(\omega_n) =  - \frac{{(i\lambda\hat{\Omega}_n    + {\xi ^ + }})}{{{\lambda^2\Omega_n ^2} + {{({\xi^ + })}^2}}}\frac{({1 + {\bf{n}}\bm{\sigma} })}{2} - \frac{({i\lambda\hat{\Omega_n}  + {\xi ^ - }})}{{{\lambda^2\Omega_n ^2} + {{({\xi^ - })}^2}}}\frac{({1 - {\bf{n}}\bm{\sigma} })}{2}\,,
\end{equation}
\end{widetext}
where $\xi^{\pm}=\xi\pm h_{\mathbf{k}}$, $\mathbf{n}=\mathbf{h}_{\mathbf{k}}/h_{\mathbf{k}}$ and the factor $\lambda$ is given by $\lambda=1+1/2\tau_{\text{sc}}\Omega_n$. One can check with this Green function that $\tau_3G^0(\mathbf{r},\mathbf{r},\omega_n)$, which enters into  the unperturbed part of self-energy Eq.(\ref{Sigma}), is given by
\begin{widetext}
\begin{equation}\label{Grr}
\tau_3G^0(\mathbf{r},\mathbf{r},\omega_n)=\sum_{\mathbf{k}}\tau_3G^0_{\mathbf{k}}(\omega_n)=
-\frac{N_F}{2}\int d\xi\left(\frac{{(i\lambda\hat{\Omega}_n    + {\xi ^ + }})}{{{\lambda^2\Omega_n ^2}+{{({\xi^ + })}^2}}}+ \frac{({i\lambda\hat{\Omega_n}  + {\xi ^ - }})}{{{\lambda^2\Omega_n ^2} + {{({\xi^ - })}^2}}}\right)\simeq -i\pi N_F\frac{\hat{\Omega}_n}{\Omega_n}\,.
\end{equation}
\end{widetext}
A substitution of this expression into Eq.(\ref{Sigma}) and  Eq.(\ref{G}) results in appearance of the factor $\lambda$ in Eq.(\ref{G0}). It should be noted that the integration over $\xi$ in Eq.(\ref{Grr}) is performed with an accuracy not higher than $h_{\mathbf{k}_F}/\mu$.

After a spatial Fourier transformation and linearization with respect to $Z$, $\bm{\nabla}\phi$ and $G_{\mathbf{k}Z}(\mathbf{q},\omega_n)$  Eq.(\ref{GZa}) takes the form
\begin{eqnarray}\label{GZa2}
&&i\left[\lambda\hat{\Omega}_n,G_{\mathbf{k}Z}\right]-\frac{1}{2}\{\mathbf{vq},G_{\mathbf{k}Z}\}-
\left[\mathbf{h}_{\mathbf{k}}\bm{\sigma},G_{\mathbf{k}Z}\right]
=\nonumber \\
&&\hat{O}[\tau_3G^0_{\mathbf{k}}]+
\frac{1}{2\tau_{\text{sc}}}\left[g,\tau_3G^0_{\mathbf{k}}\right]\,,
\end{eqnarray}
where $\mathbf{q}$ is the wave-vector and the operator $\hat{O}$ is defined by
\begin{eqnarray}\label{O}
\hat{O}[\ast]=[\hat{Z}-i\mathbf{q}\mathbf{v}\tau_3\phi_{\mathbf{q}},\ast]-\nonumber \\
(\mathbf{q}/2)
\left\{\hat{Z}-i\mathbf{q}\mathbf{v}\tau_3\phi_{\mathbf{q}}+(1/2\tau_{\text{scat}})g,\bm{\nabla}_{\mathbf{k}}\ast\right\}\,,
\end{eqnarray}
 with $\hat{Z}=\tau_3(\mathbf{Z}_{\mathbf{q}}\bm{\sigma})$.  The second term in the right-hand side of Eq.(\ref{GZa2}) describes the impurity scattering, where $g$ is given by
\begin{equation}\label{g}
g\equiv g(\mathbf{q},\omega_n)=\frac{1}{\pi N_F}\sum_{\mathbf{k}} G_{\mathbf{k}Z}(\mathbf{q},\omega_n)\,.
\end{equation}
In the basis where $\hat{\Omega}_n$ is diagonal the first term in the left-hand side of Eq.(\ref{GZa2}) can be represented as $i\lambda\Omega_n(\tau-\tau^{\prime})G_{\mathbf{k}Z\tau\tau^{\prime}}$, where $\tau=\pm 1$ are eigenvalues of $\hat{\Omega}_n/\Omega_n$ and the arguments $\mathbf{q}$ and $\omega_n$ in $G_Z$ are suppressed.  Further, it is convenient to write Eq.(\ref{GZa2}) in the form
\begin{widetext}
\begin{equation}\label{GZ2}
G_{\mathbf{k}Z\tau\tau^{\prime}}=L[G_{\mathbf{k}Z\tau\tau^{\prime}}]+R_{\mathbf{k}\tau\tau^{\prime}}\,\,\,,\,\,\, R_{\mathbf{k}\tau\tau^{\prime}}=-\frac{i}{\lambda\Omega_n(\tau-\tau^{\prime})}\left(\hat{O}(\tau_3G^0_{\mathbf{k}})+
\frac{1}{2\tau_{\text{scat}}}\left[g,\tau_3G^0_{\mathbf{k}}\right]\right)_{\tau\tau^{\prime}}\,,
\end{equation}
where $\tau\neq\tau^{\prime}$ and the operator $L$ is
\begin{equation}\label{L}
L[*]=-\frac{i}{\lambda\Omega_n(\tau-\tau^{\prime})}\left(\frac{1}{2}\{\mathbf{vq},*\}+
\left[\mathbf{h}_{\mathbf{k}}\bm{\sigma},\ast\right]\right)\,.
\end{equation}
\end{widetext}
From Eq. (\ref{GZ2}) the function $G_{\mathbf{k}Z}$ can be expressed as an expansion over the small parameter $1/\lambda \sim \tau_{\text{sc}}\Omega_n\ll 1$. We start iteration from $G_{\mathbf{k}Z}^{(0)}=R_{\mathbf{k}}$, up to $L^3$ terms, so that
\begin{equation}\label{GZ3}
G_{Z\tau\tau^{\prime}}=R_{\tau\tau^{\prime}}+L[R_{\tau\tau^{\prime}}]+L[L[R_{\tau\tau^{\prime}}]]+L[L[L[R_{\tau\tau^{\prime}}]]]\,.
\end{equation}
In this expansion the spin-dependent part of the velocity $\mathbf{v}$, which is defined below Eq.(\ref{GZa}), must be taken into account only in the linear order, since it is small by the parameter $\alpha/v_F\simeq h_{\mathbf{k}_F}/\mu$. By integrating Eq.(\ref{GZ3}) over $\mathbf{k}$ we obtain in l.h.s the density function $g$, according to Eq.(\ref{g}). This k-independent function also enters in the scattering term in r.h.s.. By this way we arrive to a set of  closed Usadel equations Eq.(\ref{Usadel}) for elements of the matrix $g$.  The integration in r.h.s. involves the unperturbed function $G^0_{\mathbf{k}}$ which, according to Eq. (\ref{GZ2}), enters into $R_{\mathbf{k}}$. In Eq.(\ref{GZ3}) it has to be integrated together with functions that vary slowly near $\mu$, as can be seen from Eqs.(\ref{L}) and (\ref{GZ2}). At the same time, $G^0_{\mathbf{k}}$ has poles that are placed close to the spin-split Fermi surfaces defined by the equations $\xi \pm h_{\mathbf{k}}=0$. With the prescribed accuracy in the integrals over $\xi$ this small splitting has to be taken into account up to the leading order with respect to $\alpha/v_F \sim h_{k_F}/\mu$. Accordingly, two typical  integrals that emerge from Eqs.(\ref{G0}) and (\ref{GZ3}) have to be  treated in the following way:
\begin{widetext}
\begin{eqnarray}\label{integrals}
\int \frac{d^2k}{(2\pi)^2} f(\mathbf{k})\left (\frac{\lambda\Omega_n\tau + \xi ^ + }{\lambda^2\Omega_n ^2 + (\xi^ +)^2} +\frac{\lambda\Omega_n\tau + \xi ^ - }{\lambda^2\Omega_n ^2 + (\xi^ -)^2}\right)&\simeq&2N_F\int d\xi \frac{d\phi}{2\pi} f(\mathbf{k}_F)\frac{\lambda\Omega_n\tau + \xi }{\lambda^2\Omega_n ^2+ \xi^2}=2\pi\tau N_F\int
\frac{d\phi}{2\pi}f(\mathbf{k}_F)\,; \nonumber \\
\int \frac{d^2k}{(2\pi)^2} f(\mathbf{k})v^i\mathbf{n}\left (\frac{\lambda\Omega_n\tau + \xi ^ + }{\lambda^2\Omega_n ^2 + (\xi^ +)^2} -\frac{\lambda\Omega_n\tau + \xi ^ - }{\lambda^2\Omega_n ^2 + (\xi^ -)^2}\right) &\simeq& -2N_F\int d\xi \frac{d\phi}{2\pi} \left [\nabla^i_{\mathbf{k}}(\mathbf{h}_{\mathbf{k}}f(\mathbf{k}))\right]_{k=k_F}
\frac{\lambda\Omega_n\tau + \xi }{\lambda^2\Omega_n ^2+ \xi^2}= \nonumber \\ -2\pi\tau N_F\int \frac{d\phi}{2\pi}\left [\nabla^i_{\mathbf{k}}(\mathbf{h}_{\mathbf{k}}f(\mathbf{k}))\right]_{k=k_F}\,,
\end{eqnarray}
\end{widetext}
where $\phi$ is a polar angle, $v^i=\nabla^i_{\mathbf{k}}\epsilon(\mathbf{k})$  and $ f(\mathbf{k})$ is an arbitrary slowly varying  function. In the second line of Eq.(\ref{integrals}) we expanded the Green's functions over $h_{\mathbf{k}}$ and used the relation $v^i\partial/\partial\xi = \nabla_{\mathbf{k}}^i$. A straightforward term-by-term analysis of integrals in Eq.(\ref{GZ3}) shows that contributions from $\mathbf{q}\bm{\nabla}_{\mathbf{k}}G^0_{\mathbf{k}}$ in the second term of Eq.(\ref{O}) for $\hat{O}[\tau_3G^0_{\mathbf{k}}]$ give only small corrections $\sim 1/\mu$ to parameters of the Usadel equation.

The current can be calculated from Eq.(\ref{j4}) where only the perturbed part $G_{\mathbf{k}Z}$ gives a nonzero result. In the leading approximation  it is sufficient to take into account only the terms up to $L^2$ in Eq.(\ref{GZ3}). By employing the integration rules (\ref{integrals}) we arrive to Eq.(\ref{jnu}) for the current.

\end{document}